\documentclass[10pt,prb,twocolumn,showpacs]{revtex4}
\usepackage{graphicx,amsmath,amssymb}
\begin{document}

\renewcommand{\Re}{\mathop{\mathrm{Re}}}
\renewcommand{\Im}{\mathop{\mathrm{Im}}}
\renewcommand{\b}[1]{\mathbf{#1}}
\renewcommand{\c}[1]{\mathcal{#1}}
\renewcommand{\u}{\uparrow}
\renewcommand{\d}{\downarrow}
\newcommand{\bsigma}{\boldsymbol{\sigma}}
\newcommand{\blambda}{\boldsymbol{\lambda}}
\newcommand{\tr}{\mathop{\mathrm{tr}}}
\newcommand{\sgn}{\mathop{\mathrm{sgn}}}
\newcommand{\sech}{\mathop{\mathrm{sech}}}
\newcommand{\diag}{\mathop{\mathrm{diag}}}
\newcommand{\half}{{\textstyle\frac{1}{2}}}
\newcommand{\sh}{{\textstyle{\frac{1}{2}}}}
\newcommand{\ish}{{\textstyle{\frac{i}{2}}}}
\newcommand{\thf}{{\textstyle{\frac{3}{2}}}}

\title{Magnetoconductance of the quantum spin Hall state}

\author{Joseph Maciejko$^{1,2}$, Xiao-Liang Qi$^{1,2}$, and Shou-Cheng Zhang$^{1,2}$}

\affiliation{$^1$Department of Physics, Stanford
University, Stanford, CA 94305, USA\\
$^2$Stanford Institute for Materials and Energy Sciences,\\
SLAC National Accelerator Laboratory, Menlo Park, CA 94025, USA}

\date\today

\begin{abstract}
We study numerically the edge magnetoconductance of a quantum spin Hall
insulator in the presence of quenched nonmagnetic disorder. For a
finite magnetic field $B$ and disorder strength $W$ on the order of
the bulk gap $E_g$, the conductance deviates from its quantized value in a manner which appears to be linear in $|B|$ at small $B$. The observed behavior is in
qualitative agreement with the cusp-like features observed in recent
magnetotransport measurements on HgTe quantum wells. We propose a
dimensional crossover scenario as a function of $W$, in which for
weak disorder $W<E_g$ the edge liquid is analogous to a disordered
spinless 1D quantum wire, while for strong disorder $W>E_g$, the
disorder causes frequent virtual transitions to the 2D bulk, where
the originally 1D edge electrons can undergo 2D diffusive motion and
2D antilocalization.
\end{abstract}


\maketitle

\section{Introduction}

A great deal of interest has been generated recently by the
theoretical prediction\cite{bernevig2006d} and experimental
observation\cite{koenig2007b,Roth2009,Buttiker2009} of the
quantum spin Hall (QSH) insulator state
\cite{kane2005a,bernevig2006a,Konig2008}. The QSH state is a novel
topological state of quantum matter which does not break
time-reversal symmetry (TRS), but has a bulk insulating gap and
gapless edge states with a distinct helical liquid property
\cite{Wu2006}. The gaplessness of the edge states is protected
against weak TRS preserving perturbations by Kramers degeneracy
\cite{Wu2006,Xu2006}. As a result, the QSH state exhibits robust
dissipationless edge transport
\cite{koenig2007b,Roth2009,Buttiker2009} in the presence of
nonmagnetic disorder.

However, in the presence of an external magnetic field which
explicitly breaks TRS, the gaplessness of the edge states is not
protected. This can be simply understood by considering the
generic form of the effective one-dimensional (1D) Hamiltonian $H$ for the QSH edge
\cite{Qi2008} to first order in the magnetic field $\b{B}$,
$H=H_0+H_1(\b{B})$, where $H_0=\hbar vk\sigma_3$ is the
Hamiltonian of the unperturbed edge, and
$H_1(\b{B})=\sum_{a=1,2,3}(\b{t}_a\cdot\b{B})\sigma_a$ is the
perturbation due to the field. $k$ is a 1D wave vector along the
edge, $v$ is the edge state velocity, $\sigma_{1,2,3}$ are the
three Pauli spin matrices, and $\b{t}_{1,2,3}$ are model-dependent
coefficient vectors\cite{Qi2008}. If $\b{B}$ points along a
special direction in space $\b{t}^*\equiv\b{t}_1\times\b{t}_2$,
then $H_1(\b{B})\propto\sigma_3$ commutes with $H_0$, the wave
vector $k$ is simply shifted, and the edge remains gapless, unless
mesoscopic quantum confinement effects become important
\cite{tkachov2010}. If $\b{B}\nparallel\b{t}^*$, then
$[H_0,H_1(\b{B})]\neq 0$ and a gap $E_\mathrm{gap}\propto|B|$
opens in the edge state dispersion.

Experimentally\cite{koenig2007b,MarkusThesis}, one observes that
the conductance $G(B)$ of a QSH device exhibits a sharp cusp-like
peak at $B=0$, and $G$ decreases for increasing $|B|$. Although
the explanation of a thermally activated behavior $G(B)\propto
e^{-E_\mathrm{gap}(|B|)/k_BT}$ with $T$ the temperature can
account qualitatively for the observed cusp, it does so only if
the chemical potential $\mu$ lies inside the edge gap which,
according to theoretical estimates\cite{Konig2008}, is rather
small ($E_\mathrm{gap}\sim 1$~meV). Experimentally, a sharp peak
is observed\cite{MarkusThesis} throughout the bulk gap ($E_g\sim
40$~meV). Furthermore, this explanation ignores the effects of
disorder. In the absence of TRS, the QSH edge liquid is
topologically equivalent to a spinless 1D quantum wire, and is
thus expected to be strongly affected by disorder due to Anderson
localization. Although the effect of disorder on transport in the
QSH state has been the subject of several recent studies
\cite{Wu2006,Xu2006,Sheng2006,Onoda2007,Obuse2007,Li2009a,Li2009b},
except for studies addressing the effect of magnetic impurities
\cite{Wu2006,Maciejko2009} there have been no theoretical
investigations of the combined effect of disorder and TRS breaking
on edge transport in the QSH state.

In this work, we study numerically the edge magnetoconductance $G$
of a QSH insulator in the presence of quenched nonmagnetic
disorder. Our main findings are: (1) For a finite magnetic field
$B$ and disorder strength $W$ on the order of the bulk energy gap
$E_g$, $G$ deviates from its quantized value $G(0)=2e^2/h$ at zero field~\cite{koenig2007b} by an amount $\Delta G(B)\equiv G(B)-G(0)$ which seems roughly linear in $|B|$ at small $B$, at least in the range of fields we study. We observe this behavior for $\mu$ across the bulk gap (Fig.~\ref{fig:ttmag_Ny}c), which agrees qualitatively with the
cusp-like features reported in Ref.~\onlinecite{koenig2007b}. (2)
The slope $\partial G/\partial B$ of $G(B)$ at small $B$ steepens rapidly when $W>E_g$ (Fig.~\ref{fig:ttmag_loc}b), which suggests that bulk states play
an important role in the backscattering of the edge states. (3)
$G$ is unaffected by an orbital magnetic field in the absence of
inversion symmetry breaking terms (Fig.~\ref{fig:ttmag_bia}a). In
the absence of such terms, $\b{t}_1$ and $\b{t}_2$ are entirely in
the $xy$ plane of the device\cite{Konig2008}, hence
$\b{t}^*\propto\hat{\b{z}}$ is out-of-plane and a perpendicular
field $\b{B}\parallel\b{t}^*$ cannot lead to backscattering, as
discussed earlier. In the presence of inversion symmetry breaking
terms, the effective edge Hamiltonian becomes $H'=\hbar
vk\sigma_3'+\sum_{a=1,2,3}(\b{t}'_a\cdot\b{B})\sigma_a'$, where
$\sigma_3'$ has nonzero components along the $1$ and $2$
directions. Then $\b{t}^{\prime *}=\b{t}'_1\times\b{t}'_2$ is not
along $\hat{\b{z}}$ anymore, and a perpendicular field
$\b{B}=B\hat{\b{z}}$ can lead to backscattering.

\section{Theoretical Model}

We start from a simple 4-band
continuum model Hamiltonian\cite{bernevig2006d,Konig2008} used to
describe the physics of the QSH state in HgTe quantum wells (QW),
\begin{equation}\label{kpH}
\c{H}(\b{k})=\left(
\begin{array}{cc}
H(\b{k}) & \Delta(\b{k}) \\
\Delta^\dag(\b{k}) & H^*(-\b{k})
\end{array}
\right),
\end{equation}
written in the $(E1^+,H1^+,E1^-,H1^-)$ basis where $E1,H1$ are the
relevant QW subbands close to the Fermi energy and $\pm$ denotes
time-reversed partners. The diagonal blocks $H(\b{k}),H^*(-\b{k})$ with
$H(\b{k})=\epsilon(\b{k})+v\b{k}\cdot\bsigma+M(\b{k})\sigma_z$ are
related by TRS and correspond to decoupled 2D Dirac-like
Hamiltonians, where $\b{k}=(k_x,k_y)$,
$\bsigma=(\sigma_x,\sigma_y)$ is a vector of Pauli matrices, and
the velocity $v$ is obtained from $\b{k}\cdot\b{p}$ theory. We
also define a quadratic kinetic energy term
$\epsilon(\b{k})=C-D\b{k}^2$ and the Dirac mass term
$M(\b{k})=M-B\b{k}^2$, where $C,D,M,B$ are $\b{k}\cdot\b{p}$
parameters. The off-diagonal block $\Delta(\b{k})$ is given by\cite{Chaoxing}
\begin{equation}\label{BIA}
\Delta(\b{k})=\left(
\begin{array}{cc}
\Delta_e k_+ & -\Delta_z \\
\Delta_z & \Delta_h k_-
\end{array}
\right),
\end{equation}
where $\Delta_e,\Delta_h,\Delta_z$ are $\b{k}\cdot\b{p}$
parameters and $k_\pm=k_x\pm ik_y$. It originates from the bulk
inversion asymmetry (BIA) of the underlying microscopic zincblende
structure of HgTe and CdTe\cite{winklerbook}. A nearest-neighbor
tight-binding (TB) model on the square lattice can be derived from
Eq. (\ref{kpH}),
\begin{equation}\label{TBmodel}
\mathcal{H}=\sum_i c^\dag_i V c_i+\sum_i\left(c_i^\dag
T_{\hat{x}}c_{i+\hat{x}}+c_i^\dag
T_{\hat{y}}c_{i+\hat{y}}+\mathrm{h.c.}\right),
\end{equation}
where the $4\times 4$ matrices $V,T_{\hat{x}},T_{\hat{y}}$ depend
solely on the $\b{k}\cdot\b{p}$ parameters introduced above.

Equations (\ref{kpH}) and (\ref{TBmodel}) correspond to a
translationally invariant system in the absence of magnetic field
or disorder. In the presence of disorder and an external magnetic
field $\b{B}=(B_x,B_y,B_z)$, we perform the substitutions
\begin{eqnarray*}
V&\longrightarrow&V+H_{Z\parallel}+H_{Z\perp}+W_i,\\
T_{\hat{x}}&\longrightarrow&T_{\hat{x}}\exp\left(\frac{2\pi i}
{\phi_0}\int_i^{i+\hat{x}}d\boldsymbol{\ell}\cdot\b{A}\right)
=T_{\hat{x}}e^{-2\pi in_zy/a},
\end{eqnarray*}
where $W_i$ is a Gaussian random on-site potential with standard
deviation $W$ mimicking quenched disorder, $\b{A}=(-B_zy,0)$ is
the in-plane electromagnetic vector potential in the Landau gauge,
$\phi_0=h/e$ is the flux quantum, and $n_z=B_za^2/\phi_0$ is the
number of flux quanta per plaquette with $a$ the lattice constant.
We use $a=30$~\AA~which is a good approximation to the continuum
limit. The in-plane Zeeman term $H_{Z\parallel}$ is given by
\cite{Konig2008}
\begin{equation}\label{HZparallel}
H_{Z\parallel}=g_\parallel \mu_B\left(
\begin{array}{cccc}
0 & 0 & B_- & 0 \\
0 & 0 & 0 & 0 \\
B_+ & 0 & 0 & 0 \\
0 & 0 & 0 & 0
\end{array}
\right),
\end{equation}
where $B_\pm=B_x\pm iB_y$, $\mu_B$ is the Bohr magneton, and the
in-plane $g$-factor $g_\parallel$ is obtained from
$\b{k}\cdot\b{p}$ calculations\cite{Chaoxing}. The out-of-plane
Zeeman term $H_{Z\perp}$ is given by\cite{Konig2008}
\begin{equation}\label{HZperp}
H_{Z\perp}=\mu_B B_z\diag
\bigl(g_{E\perp},g_{H\perp},-g_{E\perp},-g_{H\perp}\bigr),
\end{equation}
and the out-of-plane $g$-factors $g_{E\perp},g_{H\perp}$ are also
obtained from $\b{k}\cdot\b{p}$ calculations\cite{Chaoxing}. The
$\b{k}\cdot\b{p}$ parameters used in the present work correspond
to a HgTe QW thickness of $d=80$~\AA.

\begin{figure}[t]
\begin{center}
\includegraphics[width=3.5in]{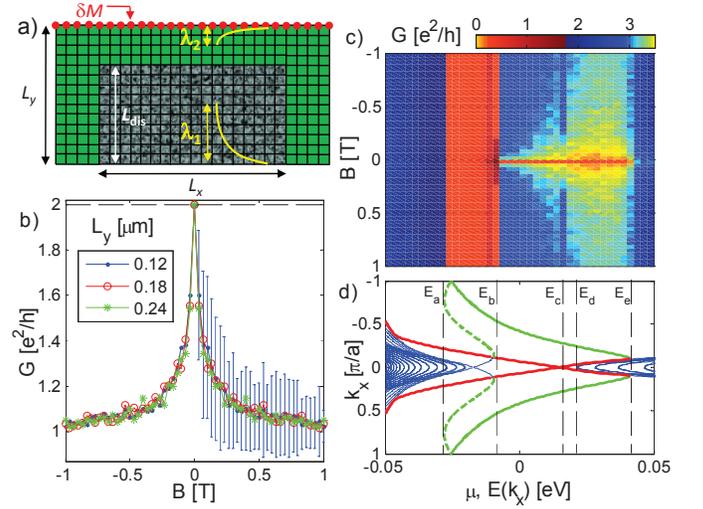}
% a) averaged over 100 impurities; b) averaged over 40 impurities
\end{center}
\caption{Magnetoconductance $G$ of a QSH edge: a) TB model with
asymmetric edge states $\lambda_2\ll\lambda_1$ to study a single disordered edge;
b) dependence of $G$ on sample width $L_y$ for disorder strength
$W=55$ meV larger than the bulk gap, length $L_x=2.4\,\mu$m,
fixed clean width $L_y-L_\mathrm{dis}=0.03\,\mu$m, and local mass term
$\delta M=-70$ meV, with error bars (plotted for
$L_y=0.12\,\mu$m and $B>0$ only) corresponding to conductance fluctuations $\delta G$;
c) dependence of $G$ on chemical potential $\mu$; d) quasi-1D spectrum of the
device illustrated in a) for zero $W,B$, showing bulk states (blue), top edge states
(green) and bottom edge states (red).}
\label{fig:ttmag_Ny}
\end{figure}

We calculate numerically the $T=0$ disordered-averaged two-terminal
conductance $G$ and conductance fluctuations $\delta G$ of a finite
QSH strip (Fig.~\ref{fig:ttmag_Ny}a) using the standard TB Green
function approach\cite{FerryGoodnick}.  We find that $N_\mathrm{dis}\sim 100$ disorder configurations are enough to achieve good convergence for
$G$ and $\delta G$. For a strip of width $L_y$
comparable to the edge state penetration depth $\lambda$, interedge
tunneling\cite{Zhou2008} backscatters the edge states even at $B=0$
and the system is analogous to a topologically trivial quasi-1D quantum
wire. To ensure that we are studying effects intrinsic to the
topologically nontrivial QSH helical edge liquid, we first need to
suppress interedge tunneling. The naive way to achieve this is to
use a very large $L_y$; however, this can be computationally rather
costly. We use a geometry (Fig.~\ref{fig:ttmag_Ny}a) which allows us
to effectively circumvent this problem while keeping $L_y$
reasonable. By adding a local Dirac mass term\cite{Konig2008}
$\delta M<0$ on the first horizontal chain of our TB model
(Fig.~\ref{fig:ttmag_Ny}a, red dots), the penetration depth
$\lambda_2$ at the top edge becomes much smaller than that at the
bottom edge $\lambda_1\gg\lambda_2$. We then add disorder only to
the last $L_\mathrm{dis}/a$ chains of the central region with
$L_\mathrm{dis}\gg\lambda_1$ and $L_y-L_\mathrm{dis}\gg\lambda_2$.
The resulting top edge states are very narrow, contribute an
uninteresting background quantized conductance independent of $B$
and $W$, and are essentially decoupled from the bottom edge states
(whose magnetoconductance we wish to study) that are effectively
propagating in a semi-infinite disordered medium.

\section{Numerical Results}

For $\mu$ inside the bulk gap, we
expect edge transport to dominate the physics. The typical
behavior of the magnetoconductance $G(B)$ for
$\b{B}=B\hat{\mathbf{z}}$ and disorder strength $W$ larger than
the bulk gap $E_g\simeq 40$~meV is shown in
Fig.~\ref{fig:ttmag_Ny}b. The cusp-like feature at $B=0$ agrees
qualitatively with the results of Ref.~\onlinecite{koenig2007b}.
$G(B)$ is independent of $L_y$, which suggests that transport is
indeed carried by the edge states. $G(B=0)$ is quantized to
$G_0\equiv 2e^2/h$ independent of $W$ up to $W=71$~meV with extremely
small conductance fluctuations $\delta G(B=0)/G_0\sim 10^{-5}$,
which confirms that interedge tunneling is negligible even for
strong disorder. Furthermore, $G$ tends to $G_0/2$ for large
$|B|\sim 1$~T, which indicates that the disordered bottom edge is
completely localized for large $W$ and $|B|$, and only the
unperturbed top edge conducts. For $W< E_g$, $G$ is approximately
quadratic in $B$ (not shown), and $|G(B)-G_0|/G_0\ll 1$ even for
large $|B|\sim 1$~T. For $B\neq 0$, we observe that the amplitude
of the fluctuations $\delta G$ does not decrease upon increasing
$N_\mathrm{dis}$, and is roughly independent of $W$ with $\delta
G/G_0\sim\c{O}(10^{-1})$ for large enough disorder $W\gtrsim E_g$.
Since in the absence of TRS the QSH system is a trivial insulator
and the edge becomes analogous to an ordinary spinless 1D quantum
wire with no topological protection, we conclude that $\delta G$
corresponds to the well-known universal conductance fluctuations
\cite{FerryGoodnick}.

The dependence of $G(B)$ on $\mu$ is plotted in
Fig.~\ref{fig:ttmag_Ny}c. We consider $W=55$~meV slightly larger
than $E_g$ (Fig.~\ref{fig:ttmag_Ny}d). This is not unreasonable as
the bulk mobility $\mu^*$ of the HgTe QW in
Ref.~\onlinecite{koenig2007b} is estimated as $\mu^*\simeq
10^5$~cm$^2$/(V$\cdot$s), which corresponds to a momentum
relaxation time $\tau=\mu^*m^*/e\simeq 0.57$~ps. The bulk carriers
at the bottom of the conduction subband have an effective mass
$m^*\simeq0.01m_e$ where $m_e$ is the bare electron mass. $\tau$
is given by $\hbar/\tau\simeq 2\pi\nu(Wa)^2$, with $\nu$ the bulk
continuum density of states at the Fermi energy given by
$\nu\simeq m^*/\pi\hbar^2$. This yields $W\simeq 22$~meV. However,
this estimate considers only bulk disorder and we expect edge
roughness to yield a higher effective $W$ on the edge.
Furthermore, this estimate is perturbative in $W$ and neglects
interband effects which are expected to occur for $W\sim E_g$. For
the chosen value of $W$ we observe that the bulk states
(Fig.~\ref{fig:ttmag_Ny}d, blue lines) are strongly localized with
$G\ll G_0$ for $\mu<E_a$ and $\mu>E_e$ in the bulk bands, while
the cusp-like feature at $B=0$ with $G(B=0)=G_0$ remains prominent
for $E_b<\mu<E_d$ in the bulk gap and even at the bottom of the
conduction band $E_d<\mu<E_e$ where the top edge states
(Fig.~\ref{fig:ttmag_Ny}d, red lines) coexist with the bulk
states. The sudden dip in $G(B\neq 0)$ for $\mu\sim E_c\simeq
15$~meV corresponds to the opening of the small edge gap discussed
earlier. Finally, $G\simeq G_0$ is almost independent of $B$ for
$E_a<\mu<E_b$, where the disordered bottom edge and bulk states
are mostly localized while the clean top edge supports another
channel (Fig.~\ref{fig:ttmag_Ny}d, dashed green line), with a
total top edge conductance of $G=G_0$.

\begin{figure}
\begin{center}
\includegraphics[width=3.5in]{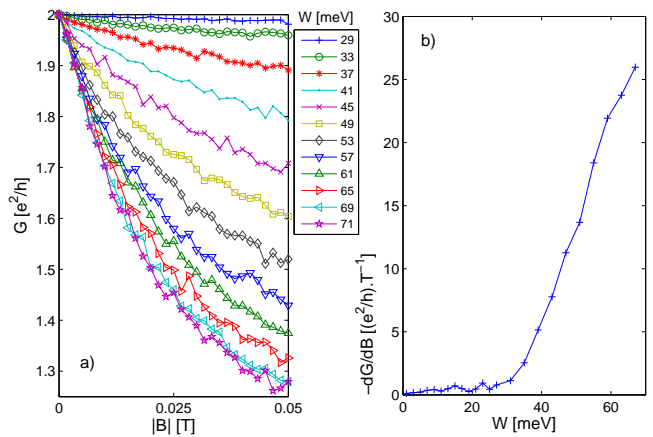}
% averaged over ~1000 impurities
\end{center}
\caption{a) Magnetoconductance for various disorder strengths $W$;
b) small-$B$ slope of the magnetoconductance (obtained by linear regression for $0<B<15$~mT).
Device size is $(L_x\times L_y)=(2.4\times 0.12)\,\mu$m$^2$.}
\label{fig:ttmag_loc}
\end{figure}

The magnetoconductance for $B=B_z$ and various values of $W$ is plotted
in Fig.~\ref{fig:ttmag_loc}. Although not evident from the figure, $G(B)$ is approximately quadratic in $B$ for
$W<E_g$, and approximately linear in $|B|$ at small $B$ for $W>E_g$ (Fig.~\ref{fig:ttmag_loc}a). The slope of $G(B)$ at small fields (obtained by linear regression for $0<B<15$~mT where the dependence is approximately linear) is plotted in Fig.~\ref{fig:ttmag_loc}b, and is seen to increase rapidly
for $W\gtrsim E_g\simeq 40$~meV. For $B=0$, we have essentially $G=G_0$ independent of $W$ (Fig.~\ref{fig:ttmag_loc}a). This contrasts
with the results of Ref.~\onlinecite{Sheng2006,Li2009a} where
deviations from $G=G_0$ at $B=0$ occur for $W$ larger than some
critical value $W_c>E_g$. The reason for this difference is that
in Ref.~\onlinecite{Sheng2006,Li2009a}, disorder-induced collapse
of the bulk gap is accompanied by the edge states penetrating
deeper into the bulk and eventually reaching the opposite edge,
such that interedge tunneling takes place and causes
backscattering. Here, due to our special geometry
(Fig.~\ref{fig:ttmag_Ny}a) the top edge state is unperturbed and
always remains localized near the edge, out of reach of the bottom
edge state, even as the latter penetrates deeper into the
disordered bulk for increasing $W$.

\begin{figure}
\begin{center}
\includegraphics[width=3.5in]{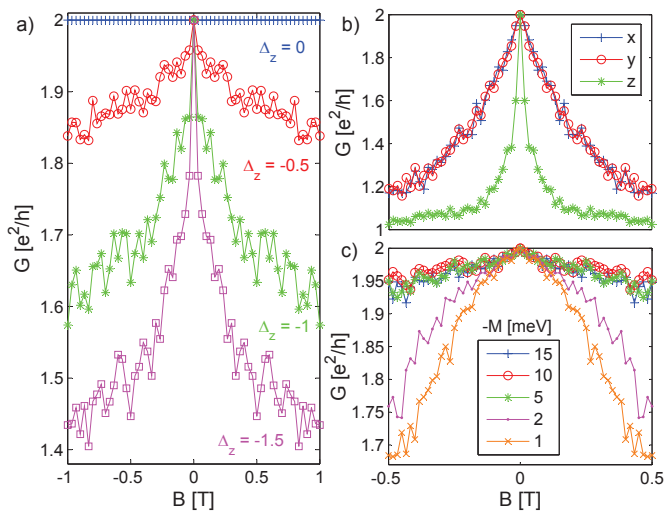}
% averaged over 100 impurities, to make it look smoother
% (however amplitude is the same with 40 impurities)
\end{center}
\caption{Dependence of the magnetoconductance $G$ on
a) strength of the $\b{k}$-independent BIA
term $\Delta_z$ with $\Delta_e=\Delta_h=0$; b) magnetic field
orientation; c) Dirac mass term $M<0$. Sample size is
$(L_x\times L_y)=(2.4\times 0.12)\,\mu$m$^2$, disorder strength
is $W=55$ meV for a),b) and $W=30$ meV for c).}\label{fig:ttmag_bia}
\end{figure}

The BIA term $\Delta_\b{k}$ has an important effect on $G$ for
$B=B_z$ (Fig.~\ref{fig:ttmag_bia}a). For simplicity, we set
$\Delta_e=\Delta_h=0$ and consider only the effect of $\Delta_z$.
For $\Delta_z=0$, the perturbation $\mathcal{H}'=e\b{j}\cdot\b{A}$
due to an orbital field, with $e$ the electron charge and $\b{j}$
the current operator, has no matrix element between the spin
states of a counterpropagating Kramers pair on a given edge
\cite{Konig2008}, and $G$ is unaffected. For an in-plane field,
$H_{Z\parallel}$ does have a nonzero matrix element between these
states, and there is a nontrivial magnetoconductance even in the
absence of BIA.

The dependence of $G(B)$ on the orientation of $\b{B}$ is plotted
in Fig.~\ref{fig:ttmag_bia}b. The $g$-factors\cite{Chaoxing} used
in the Zeeman terms are such that the Zeeman energies for in-plane
and out-of-plane fields are of the same order\cite{Konig2008}.
The in-plane vs out-of-plane anisotropy
(Fig.~\ref{fig:ttmag_bia}b, $x,y$ vs $z$) arises from the orbital
effect of the out-of-plane field $B=B_z$, which is absent for an
in-plane field. In our model, the in-plane anisotropy is very weak
(somewhat visible on Fig.~\ref{fig:ttmag_bia}b for $|B|\sim 1$~T),
and is due to the inequivalence between the transport $x$ and
confinement $y$ directions. Finally, the $B=0$ peak in $G$ is more
pronounced for a smaller mass term $M$\cite{Konig2008} in the
Dirac Hamiltonians $H_\b{k},H^*_{-\b{k}}$
(Fig.~\ref{fig:ttmag_bia}c). Since $E_g\propto|M|$ approximately,
a smaller $|M|$ results in a larger dimensionless disorder
strength $W/E_g$, which is equivalent to an increase in $W$ (see
Fig. \ref{fig:ttmag_loc}b).

Although the mechanism behind the observed negative
magnetoconductance $\Delta G\propto-|B|$
(Fig.~\ref{fig:ttmag_Ny},\ref{fig:ttmag_loc}) for an orbital field
$B=B_z$ cannot be unambiguously inferred from our numerical
results, a dependence linear in $|B|$ for small
$B$ and the requirement of `strong' disorder $W\gtrsim E_g$ for
its observation seem to indicate that the effect has a
nonperturbative character. A treatment which is perturbative in
$W$ and $B$ yields at most, to leading order, the result
$-\Delta G\propto\ell^{-1}\propto W_\mathrm{eff}^2(B)\propto B^2$,
where $\ell$ is the mean free path\cite{Lee1985} and
$W_\mathrm{eff}(B)\equiv W|B|/B_0$ is some effective disorder
strength, with $B_0^{-1}\propto\Delta_z$ if only the effect of
$\Delta_z$ is considered for simplicity. For `weak' disorder $W<
E_g$, the 1D edge states which enclose a negligible amount of flux
are the only low-energy degrees of freedom, and the magnetic field
only has a perturbative effect on them. Indeed, if we choose the
gauge $\b{A}=\bigl(B_z(L_y-y),0\bigr)$, for sufficiently small
$B_z$ we have that $\b{A}$ is small for $L_y-\lambda_1\lesssim
y<L_y$ with $\lambda_1\ll L_y$ where the bottom edge state
wavefunction has finite support (Fig.~\ref{fig:ttmag_Ny}a), and
the effect of an orbital field $B_z$ on a single edge can be
treated perturbatively. In this case, the amplitude $\propto
W_\mathrm{eff}(B)$ in perturbation theory for a leading order
backscattering process on a single edge involves one power of
$\Delta_z$ and one power of $B_z$ to couple the spin states of the
counterpropagating Kramers partners\cite{Konig2008} (with no
momentum transfer as our choice of gauge preserves translational
symmetry in the $x$ direction), and one power of $W$ to provide
the necessary momentum transfer for backscattering. Our
observation that $\Delta G\propto-B^2$ for $W<E_g$ corroborates
this physical picture. On the other hand, the cusp-like feature at
$B=0$ (Fig.~\ref{fig:ttmag_Ny}b) occurs for `strong' disorder
$W\gtrsim E_g$, which seems to indicate that the bulk states play
an important role. This leads us to a different physical picture.
For $W\gtrsim E_g$, the edge electrons easily undergo virtual
transitions to the bulk. In other words, the emergent low-energy
excitations for $W\gtrsim E_g$ extend deeper into the bulk than
the `bare' edge electrons. The electrons spend a significant
amount of time diffusing randomly in the bulk away from the edge,
with their trajectories enclosing finite amounts of flux before
returning to the edge, which endows the orbital field with a
nonperturbative effect. In this way the conventional picture of 2D
antilocalization (AL)\cite{Bergmann1984} can apply, at least
qualitatively, to a single disordered QSH edge. We are thus led to
the interesting picture, peculiar to the QSH state, of a
dimensional crossover between 1D AL
\cite{Zirnbauer1992,Takane2004} in the weak disorder regime
$W<E_g$ with the orbital field having a perturbative effect, to an
effect analogous to 2D AL in the strong disorder regime $W>E_g$ with
the orbital field having a nonperturbative effect.

\section{Conclusion}

We have shown that `strong' disorder effects
$W/E_g\sim 1$ in a QSH insulator in the presence of a magnetic
field $B$ and inversion symmetry breaking terms can give rise to a cusp-like feature in the two-terminal edge
magnetoconductance with an approximate linear dependence $\Delta G(B)\propto -|B|$ for
small $B$. These results are in good qualitative agreement with experiments. A possible physical intepretation of our results
consists of a dimensional crossover scenario where a weakly
disordered, effectively spinless 1D edge liquid crosses over, for
strong enough disorder, to a state where disorder enables frequent
excursions of the edge electrons into the disordered flux-threaded
2D bulk, resulting in a behavior reminiscent of 2D AL.

We wish to thank M. K\"{o}nig, H. Buhmann, L. W. Molenkamp, E. M.
Hankiewicz, C. X. Liu, T. L. Hughes, R. D. Li, H. Yao, and M.
Bourbonniere for insightful discussions. This work was supported
by the Department of Energy, Office of Basic Energy Sciences,
Division of Materials Sciences and Engineering, under contract
DE-AC02-76SF00515. JM acknowledges support from the National Science and Engineering Research Council of Canada, the Fonds Qu\'{e}b\'{e}cois de la Recherche sur la Nature et les Technologies, and the Stanford Graduate Fellowship program. Computational work was made
possible by the computational resources of the Stanford Institute
for Materials and Energy Science, and those of the Shared
Hierarchical Academic Research Computing Network
(www.sharcnet.ca).

\bibliography{mag}

\end{document}